\documentclass[twocolumn,showpacs,prl]{revtex4}

\usepackage{graphicx}

\usepackage{amssymb}

\usepackage[center]{subfigure}
\begin{document}

\title{Extreme fluctuations and the finite lifetime of the turbulent state}
\author{Nigel Goldenfeld$^1$, Nicholas Guttenberg$^1$ and Gustavo Gioia$^2$}
\affiliation{$^1$Department of Physics, University of Illinois at
Urbana-Champaign, 1110 West Green Street, Urbana, Illinois, 61801-3080}
\affiliation{$^2$Department of Mechanical Science and Engineering,
University of Illinois, 1206 West Green Street, Urbana, IL 61801-3080}

\begin{abstract}
We argue that the transition to turbulence is controlled by large
amplitude events that follow extreme distribution theory.  The theory
suggests an explanation for recent observations of the turbulent state
lifetime which exhibit super-exponential scaling behaviour with
Reynolds number.

\end{abstract}
\maketitle

\def\Re{\textrm{Re}}

The fundamental nature and stability of the turbulent state of fluids
remains an open and challenging question.  Fluid flow is characterized
by a dimensionless number $\Re$, which depends on the characteristic
length $L$, velocity $U$ and kinematic viscosity of the fluid $\nu$
through the relation $\Re\equiv UL/\nu$.  As the Reynolds number
increases from zero, the flow becomes increasingly structured and
eventually statistical in nature, and at large $\Re$, the flow is
said to be turbulent\cite{REYN83}. The conventional assumption---that
the turbulent state is absolutely stable---has been challenged recently
by a series of theoretical\cite{lagha2007mtp} and experimental
probes\cite{daviaud1992stt,bottin1998sat,hof2004eon,HOF08b,borrero2009transient}
of the transition to turbulence.  Taken as a whole, these works suggest
that turbulence might, in some flow regimes at least, be a long-lived
metastable
state\cite{hof2006flt,lagha2007mtp,eckhardt2007ttp,eckhardt2008ttp,eckhardt2008dsa}.
Such a view would be consistent with the fact that long-lived transient
turbulent states can be excited as finite-amplitude instabilities of
the laminar state, so that the laminar and turbulent states can coexist
(for a review of foundational work in this area, see e.g. Ref.
\cite{grossmann2000osf}; recent developments are summarized in Refs.
\cite{KERS05,eckhardt2007ttp,eckhardt2008ttp,eckhardt2008dsa}).  However, the
question remains as to whether the turbulent state is ever sustainable
with an infinite lifetime for finite Reynolds numbers.  This is a
difficult experimental question to decide, because the lifetime of the
turbulent state can become so long that measurements become
impossible.  With the necessary restriction to a small range of
Reynolds numbers, the data have, until recently, been difficult to
interpret in a compelling way.

In a set of elegant and remarkably accurate experiments on transitional
pipe turbulence\cite{HOF08b},  Hof et al. have brought into question
the idea that pipe flow turbulence is stable at long times beyond a
finite critical Reynolds number\cite{FAIS04,WILL07}. The laminar state
of a straight smooth pipe flow is linearly stable at all Reynolds
numbers (see e.g. Ref. (\cite{meseguer2003lpf})), but a sufficiently
large perturbation triggers localized turbulent puffs that persist for
long times. The decay of the transient turbulent state is reported to
follow a Poisson distribution, with a lifetime $\tau(\Re)$ that
increases sharply with increasing Reynolds number.  The new
measurements of the lifetime of these localized puffs\cite{HOF08b}
reveal that $\tau(\Re)$ apparently only diverges at infinite Reynolds
number, scaling in a super-exponential way with $\Re$. Similar
observations in another linearly stable flow---Taylor-Couette flow with
outer cylinder rotation---have recently been reported by
Borrero-Echeverry et al.\cite{borrero2009transient}.

In this article, we show that the form of the experimental data is
consistent with a simple and general interpretation predicated on the
use of extremal statistics. Our approach is related to the notion that
the transient turbulent phenomena reflect escape from a low-dimensional
dynamical attractor\cite{crutchfield1988art,kantz1985rsa,tel2008cts},
but we conceive turbulence as a spatially-extended phenomenon with a
large number of degrees of freedom.  The determining factor for the
suppression of a puff is the probability that the largest fluctuation
in a spatio-temporal interval consisting of multiple fluctuations fails
to attain a threshold value. Thus we need to calculate the probability
that the maximum amplitude of turbulent velocity fluctuations $\delta v
(\vec{x},t)$ falls below some threshold value, which we term $B_c$. We
will assume below that once the turbulence has been sufficiently
suppressed, the turbulent state is quenched, an assumption
consistent with previously published
analyses\cite{willis2008turbulent,schneider2008lifetime,MANN09}.  Our
calculation shows that the super-exponential dependence of the lifetime
of the turbulent state is a generic result of extremal statistics.

In order to understand the lifetime of turbulent puffs,
we assume that turbulent velocity configurations may be regarded as
independent beyond a correlation time $\tau_0$, and that there is a
probability $p$ that the puff will be suppressed within each time
interval $\tau_0$. Then, the lifetime statistics will be Poisson. The
probability $P$ that turbulence persists to a time $t$ after becoming
established at a time $t_0$ is  $P=(1-p)^M$, where the number of
intervals is $M= (t-t_0)/\tau_0$. Therefore
\begin{equation}
\ln(P)= M \ln(1-p) = \frac{1}{\tau_0} (t-t_0) \ln(1-p),
\end{equation}
and so it follows that $\tau_0/\tau = - \ln(1-p)$. Since $1\gg
p>0$, we can estimate $\ln (1-p) = -p$ and therefore express the lifetime in the form

\begin{equation}
\tau = \tau_0/ p
\end{equation}
where $p$ depends on Re.

We now determine how $p$ depends on the Reynolds number of the flow and
potentially other factors.  Within a spatial and temporal interval,
multiple fluctuations occur, sampled from the turbulent velocity
distribution $P_T(\delta v)$. The  energy associated with these
fluctuations is proportional to $\delta v^2$.  We assume that when the
energy fails to attain a certain
threshold\cite{willis2008turbulent,schneider2008lifetime,MANN09} at all
points in the puff, the turbulent state becomes unstable and decays.
Thus if the largest velocity fluctuation is less than the threshold,
all of the turbulent fluctuations are less than the threshold.
Accordingly, it is necessary to calculate the probability distribution
of the maximum of the fluctuations within a puff, in order to ensure
that the largest fluctuation is below the threshold.

We consider primarily a Gaussian distribution of velocity fluctuations,
but our arguments also apply to the case of an exponential distribution
(as is the case at high Reynolds numbers).  We seek the probability
distribution $P_M (x)$ for the maximum $x$ of a set of energy
fluctuations $\{\delta v_i^2\}$, where $i=1\dots N$. $N$ represents the
number of degrees of freedom, and should scale with the size of the
turbulent puff, denoted here by $\lambda$. Standard results from
extreme statistics theory show that the appropriate result is the
family of Fisher-Tippett distributions\cite{fisher1928lff}.  In
particular, the universality class for $P_M$ must be the Type I
Fisher-Tippett distribution, sometimes known as the Gumbel
distribution\cite{gumb58,bouchaud1997universality}
\begin{equation}
P_M(x) = \frac{1}{\beta} \exp(-(x-\mu)/\beta) \exp(\exp(-(x-\mu)/\beta)),
\end{equation}
where $\beta$ sets the scale and $\mu$ the location of the
distribution.  Note that the scale and location will depend on $N$, because the maximum of a
set of random variables will be an increasing function of the number of random variables.  In
particular, for the Gaussian case, Fisher and Tippett showed that
asymptotically
\begin{equation}
\mu \sim\sqrt{\log N}, \qquad \beta \sim 1/\sqrt{\log N}.
\label{eqn:scale}
\end{equation}

The mean and standard
deviation of the Gumbel distribution are $\mu + \Gamma\beta$ and
$\beta\pi/\sqrt{6}$ respectively, where $\Gamma \approx 0.577$ is the
Euler-Mascheroni constant.  The corresponding cumulative distribution
is the probability that $x < X$, and is given by
\begin{equation}
F(X)\equiv \int_{-\infty}^X P_M(x)\,dx = \exp(-\exp(-(X-\mu)/\beta)).
\label{eqn:Gumbel}
\end{equation}
Thus, $p = F(B_c)$, where $B_c$ is the threshold.

We anticipate that $B_c$ is a decreasing function of $\Re$, reflecting
the intuition that at higher $\Re$, turbulence can be more easily
sustained by small fluctuations.  
We will consider the behavior of $B_c$, as this sets the threshold
in the distribution of energy maxima. The experiments are conducted in
nominally smooth pipes within a narrow range of $\Re$, so it is
appropriate to expand $B_c$ about a particular Reynolds
number $\Re_0$, leading to $B_c=B_c^0 + B_c^1 ( \Re - {\Re}_0 ) +
O(\Re^2)$, where $B_c^0$ and $B_c^1$ are coefficients. In order to
describe the same Reynolds number regime of the experiments, $\Re_0$
may be interpreted to be a characteristic Reynolds number at which
localized turbulent puffs first are observable, so that the lifetime is
order $\tau_0$. This onset is not a precisely defined point, but for
concreteness we will specifically define it to be the Reynolds number
where $\tau = e \tau_0$.  We will see that this choice simplifies the
analysis below, but we emphasize that irrespective of whether or not we
use the coefficient $e$ or some other number of O(1), our main
predictions for the super-exponential distribution are not affected.
The freedom of choice in this definition of Re$_0$ is completely
analogous to the arbitrariness in the definition of the coexistence
point between liquid and gas, which is also dominated by nucleation
phenomena, as has also been noticed by Manneville\cite{MANN09}.

Collecting results, we find that the average lifetime of a turbulent
puff will have the approximate functional form:
\begin{equation}
\tau = \tau_0 \exp(\exp(-(B_c^0+B_c^1 (\Re - \Re_0) +O((\Re-\Re_0)^2)))
\label{eqn:tau-expansion}
\end{equation}
in agreement with experimental findings. The coefficient $\tau_0$ may
in principle depend on pipe length or aspect ratio if these factors
change the spatial scale on which regions of the localized turbulent
puff are statistically independent, and the timescale on which the
state of the puff loses memory of previous states.  The Taylor
expansion only needs to be carried out to first order given the
(understandably) small range of Reynolds number over which the
experiments measuring the lifetime of turbulent puffs have been
conducted.  From
(\ref{eqn:Gumbel}) we can write
\begin{equation}
\ln\ln (\tau/\tau_0)= - (B_c-\mu)/\beta = c_1\Re + c_2
\end{equation}
where we have used the notation of Ref.\ (\cite{HOF08b}) to denote the
coefficients $c_1$ and $c_2$ of the linear fit to the data.  Comparing
with Eq.\ (\ref{eqn:tau-expansion}) we read off that
\begin{equation}
c_1 = -B_c^1/\beta \qquad c_2 = -(B_c^0 - \mu - B_c^1 \Re_0)/\beta
\label{eqn:c1c2}
\end{equation}
Now, at $\Re_0$, the lifetime becomes comparable to the correlation
time $\tau_0$, and to be concrete, we chose that for $\Re=\Re_0$,
$\tau=e\, \tau_0$, although it is straightforward to verify that our
results have only a very weak dependence on the precise coefficient
used.  Then from Eq.\ (\ref{eqn:tau-expansion}), we see that $B_c^0 =
\mu$, and thus the ratio of the coefficients $c_2/c_1 = - \Re_0$. The
physical interpretation, if any, of the fitting parameter Re$_0$ is not
clear to us, because although it is defined loosely as a characteristic
Reynolds number below which the lifetime of the turbulent state is too
small to be observable, any systematic dependence on intrinsic features
of turbulence, flow geometry and perhaps wall roughness is beyond the
scope of our work and of the experiments available at this time.
Moreover, the choice of definition of the Reynolds number in any
particular geometry is not unique once multiple length-scales are
present, as in the case of Taylor-Couette flow, and thus it is
hard to identify a unique and consistent definition of Re and
Re$_0$ in order to retain meaning across different flow geometries.

We conclude with a comment about the dependence of $\tau$ on puff
length $\lambda$.  The number of degrees of freedom $N$ active in the
turbulent puff is proportional to $\lambda$.  In our approach, the
$\lambda$-dependence of $\tau$ can be estimated by substituting the
scaling of $\mu$ and $\beta$ as given by Eq. (\ref{eqn:scale}) into our
formula for $\tau$.  This yields that $\log (\tau) \propto \lambda$ to
leading order, showing that the super-exponential scaling with Reynolds
number does not translate into a super-exponential scaling with length,
and is consistent with numerical measurements reported in Ref.
(\cite{schneider2008lifetime}). This prediction also applies if the
probability distribution for velocity fluctuations is exponential,
because in this case $\mu \sim \log N$, i.e. as $\log \lambda$, but $\beta$
does not scale with $N$\cite{fisher1928lff}.

The interpretation of the lifetime statistics given here is related to
that suggested recently by Manneville\cite{MANN09}, and earlier
considerations by Pomeau on nucleation phenomena in
hydrodynamics\cite{POME86}.  Indeed, any spatially-extended dynamical
system with a memoryless sub-critical bifurcation should be expected to
yield extremal statistics, and we hope to report on detailed
calculations of this phenomenon in a future publication\cite{SIPO2010}.

We are grateful for discussions with Michael Schatz and thank Pinaki
Chakraborty for comments on the manuscript.  We thank the referees for
constructive scepticism about an earlier version of this manuscript.
This material is based upon work supported by the National Science
Foundation under Grant No. NSF DMR 06-04435.

\bibliographystyle{apsrev}
\bibliography{turbulence-extreme}
\end{document}